\documentclass[12pt]{iopart}
\usepackage{graphicx}

\def\ital{{\it et al }}
\def\mathbi#1{\ensuremath{\textbf{\em #1}}}

\begin{document}

\title{Probing single magnon excitations in Sr$_2$IrO$_4$ using O $K$-edge resonant inelastic X-ray scattering}
\author{X Liu$^{1,2,3}$, M P M Dean$^2$, J Liu$^{4,5}$, S G Chiuzb\u{a}ian$^{6,7}$, N Jaouen$^7$, A Nicolaou$^7$, W G Yin$^2$, C Rayan Serrao$^8$, R Ramesh$^{4,5,8}$, H Ding$^{1,3}$ and J P Hill$^2$ }

\address{$^1$ Beijing National Laboratory for Condensed Matter Physics and Institute of Physics, Chinese Academy of Sciences, Beijing 100190, China}
\address{$^2$ Condensed Matter Physics and Materials Science Department, Brookhaven National Laboratory, Upton, New York 11973, USA}
\address{$^3$ Collaborative Innovation Center of Quantum Matter, Beijing, China}
\address{$^4$ Department of Physics, University of California, Berkeley, California 94720, USA}
\address{$^5$ Materials Science Division, Lawrence Berkeley National Laboratory, Berkeley, California 94720, USA}
\address{$^6$ Sorbonne Universit\'{e}s, UPMC Univ Paris 06, UMR 7614, Laboratoire de Chimie Physique-Mati\`{e}re et Rayonnement, 11 rue Pierre et Marie Curie, F-75005 Paris, France}
\address{$^7$  Synchrotron SOLEIL, Saint-Aubin, BP 48, 91192 Gif-sur-Yvette, France}
\address{$^8$  Department of Materials Science and Engineering, University of California, Berkeley, California 94720, USA}
\ead{xliu@iphy.ac.cn}

\begin{abstract}
Resonant inelastic X-ray scattering (RIXS) at the $L$-edge of transition metal elements is now commonly used to probe single magnon excitations. Here we show that single magnon excitations can also be measured with RIXS at the $K$-edge of the surrounding ligand atoms when the center heavy metal elements have strong spin-orbit coupling. This is demonstrated with oxygen $K$-edge RIXS experiments on the perovskite Sr$_2$IrO$_4$, where low energy peaks from single magnon excitations were observed. This new application of RIXS has excellent potential to be applied to a wide range of magnetic systems based on heavy elements, for which the $L$-edge RIXS energy resolutions in the hard X-ray region is usually poor.
\end{abstract}

\submitto{Journal of Physics: Condensed Matter}

\maketitle

Recently, the resonant inelastic X-ray scattering technique has achieved great progress in condensed matter physics research \cite{Ament2011}. It has now been clearly demonstrated that RIXS is not only suitable for studying charge and orbital excitations, but it is also a powerful tool to probe magnetic dynamics. With RIXS at the $L$-edge of the magnetic elements, the magnetic dispersion curves were mapped out on various transition metal compounds such as cuprates  \cite{DeanReview2015}  and iridates \cite{KimReview2012}. Although the leading term of the photon-electron interaction does not directly couple to the spin, the strong spin-orbit coupling (SOC) of the core level {\it p} states accessed at the $L$-edge makes direct spin-flip scattering possible in the RIXS process \cite{Ament2009, Braicovich2010}. Combined with the large energy transfer and appreciable momentum transfer capability of X-ray scattering, this makes RIXS an attractive tool to measure magnetic dispersions, especially in situations where small sample volumes make inelastic neutron scattering difficult to apply.

Depending on the resonant energy of the targeted element, RIXS is naturally divided into two classes: soft X-ray and hard X-ray \cite{Ament2011}. In a soft RIXS setup,  X-ray dispersing elements are diffraction gratings, which typically operate efficiently in an X-ray energy range up to about 2000 eV. The $L$-edges of the $3d$ transition metal elements all fall below this energy scale. For high atomic number, $Z$, transition metal elements, the $L$-edge resonant energies are in the hard X-ray region and the RIXS setup is totally different. Instead of continuously tunable gratings, high quality single crystals are needed as the key optical elements. The RIXS resolution crucially depends on the availability of a high quality single crystal with a Bragg diffraction peak close to back-scattering at the energy of the $L$-edge of the targeted element. This requirement severely limits the application of RIXS, and thus by far the majority of hard x-ray RIXS studies of single spin-flip magnetic excitations have focused on $5d$ iridates \cite{Kim2012, Kim2012_327, Yin2013, Kim2014}. 

To overcome this resolution limitation, we suggest that it is possible to probe single magnon excitations with RIXS at the $K$-edge of the surrounding ligand atoms whose $p$ orbitals hybridize with the $d$ orbitals of a heavy element. In this RIXS process, a $1s$ core electron of the ligand atom is excited into the outer level $p$ states and the $d$ shell electronic configuration of the transition metal element is altered through the $p$-$d$ hybridization. The key ingredient to flip a spin, namely the strong spin-orbit coupling, is fulfilled by the $d$ electrons of the high $Z$ index transition metal elements instead of the 1$s$ core level of the ligand atoms. The energy resolution at the $K$-edge of the common ligand atoms is often particularly high. For example, the best energy resolution currently achieved for RIXS at the oxygen $K$-edge in the soft X-ray region is about 50~meV \cite{LeePRL2013}, which is much better than the majority of $5d$ elements probed using $L$-edge RIXS in the hard X-ray region. Furthermore the energy resolution is set to improve significantly in coming years due to the development of new soft x-ray RIXS beamlines \cite{SIX, ERIXS, I21, VERITAS, Huang2014}. 

For demonstration, we carried out oxygen $K$-edge RIXS experiments on the perovskite Sr$_2$IrO$_4$. This material is antiferromagnetically ordered at low temperature ($T_N \approx 230$~K), and its spin-wave magnetic dispersion has been mapped out in detail using  Ir $L_3$-edge RIXS  \cite{Kim2012}. By measuring the RIXS spectrum at the oxygen $K$-edge, peaks consistent with single magnon excitations were observed.  Our results demonstrate that the oxygen $K$-edge RIXS is a powerful tool in probing the magnetic excitations in 5$d$ transition metal oxides. 

The oxygen $K$-edge RIXS experiments were performed using the AERHA instrument at the SEXTANTS beamline at the SOLEIL Synchrotron in France \cite{Chiuzbaian2014}. The Sr$_2$IrO$_4$ samples were 200~nm thick epitaxial films deposited on SrTiO$_3$ substrates using pulsed laser deposition \cite{Serrao2013}. The sample was orientated such that the tetragonal crystal axes $a\approx 3.8$ and $c\approx 25.7$~\AA{}  were in the scattering plane. All data were collected at 25~K and the RIXS spectra were taken with the detector at a fixed scattering angle of $85^\circ$.

\begin{figure}[!t]
\begin{center}
\includegraphics[width=8 cm]{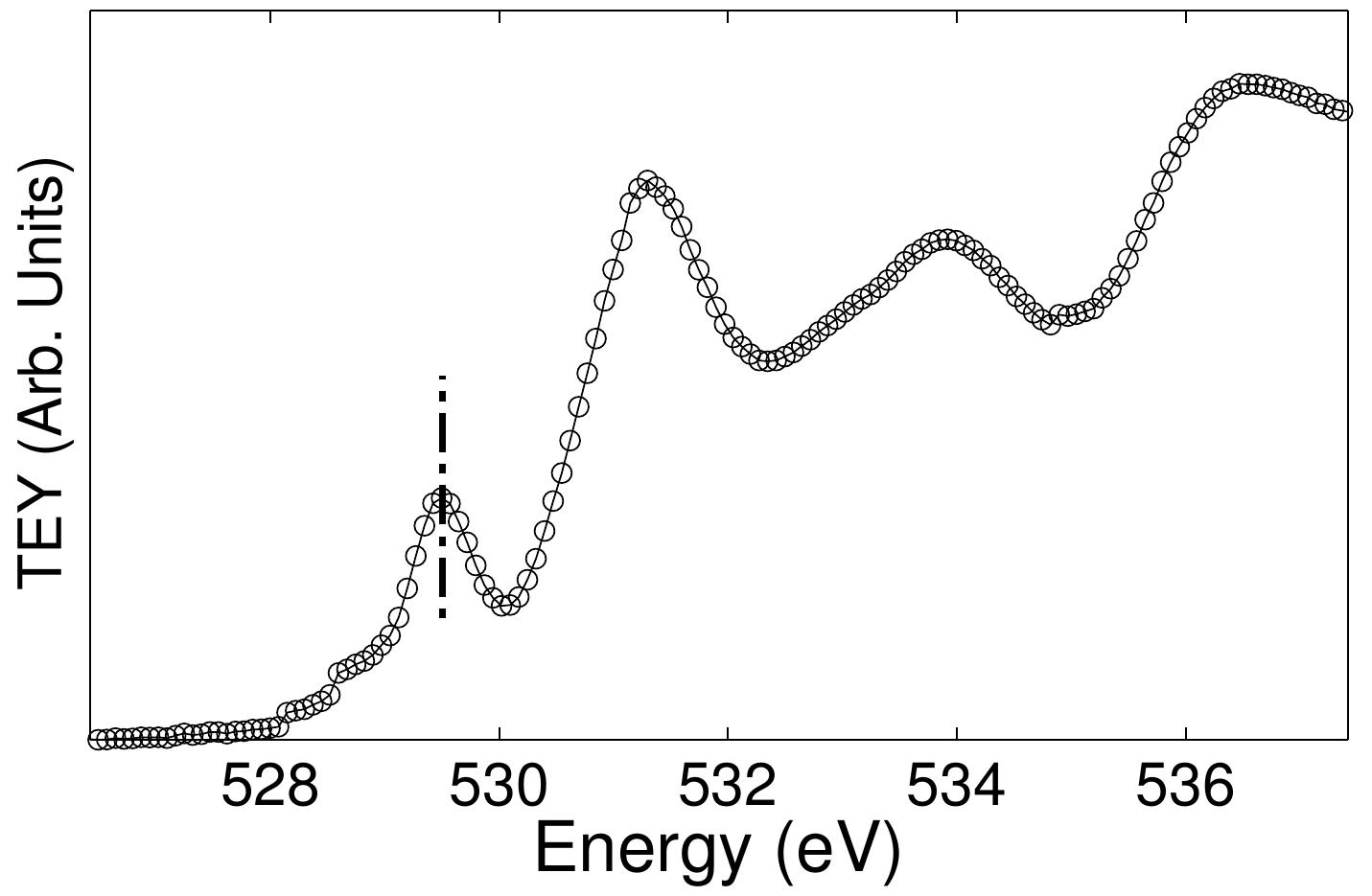}
\end{center}
\caption{Incident X-ray energy dependence of the total electron yield (TEY) intensity at the oxygen $K$-edge taken with the incident X-ray polarization vector to be $15^{\circ}$ away from surface normal. The dashed line indicates the prepeak at 529.5~eV.}
\label{fig:TEY}
\end{figure}

For different incident X-ray polarization, the X-ray couples to Ir $5d$ orbitals differently, which gives rise to different pre-peak features at the oxygen $K$-edge absorption curve \cite{Sala2014}. To maximize the cross-section, we chose the excitation into the $d_{xy}$ and $d_{yz}$ Ir orbitals via the in-plane oxygen atoms. The resonant energy of these excitations is determined by oxygen $K$-edge total electron yield spectra, shown in figure~\ref{fig:TEY}. The data were collected with the incident X-ray polarization vector to be 15 degrees away from surface normal. Thus it is mainly along the crystal $c$ direction. The pre-peak at 529.5~eV (indicated by a dashed line) is the excitation of in-plane oxygen $1s$ core electron in to the Ir $d_{xy}$ and $d_{yz}$ orbitals via $p$-$d$ hybridization.  

%\begin{figure}[h]
%\includegraphics[width=0.45\textwidth]{figure1.pdf}

With the incident X-ray energy at 529.5 eV on the prepeak at the oxygen $K$-edge, the RIXS spectra were taken with the detector at a fixed scattering angle of $85^\circ$. By rotating the sample $\theta$ angle, the total momentum transfer, \mathbi{Q}, varies in the $H$-$L$ plane as $(H, 0, L)$. Depending on the relative angle between the incident %\begin{figure}[h]
\begin{figure}[!t]
\begin{center}
\includegraphics[width=8 cm]{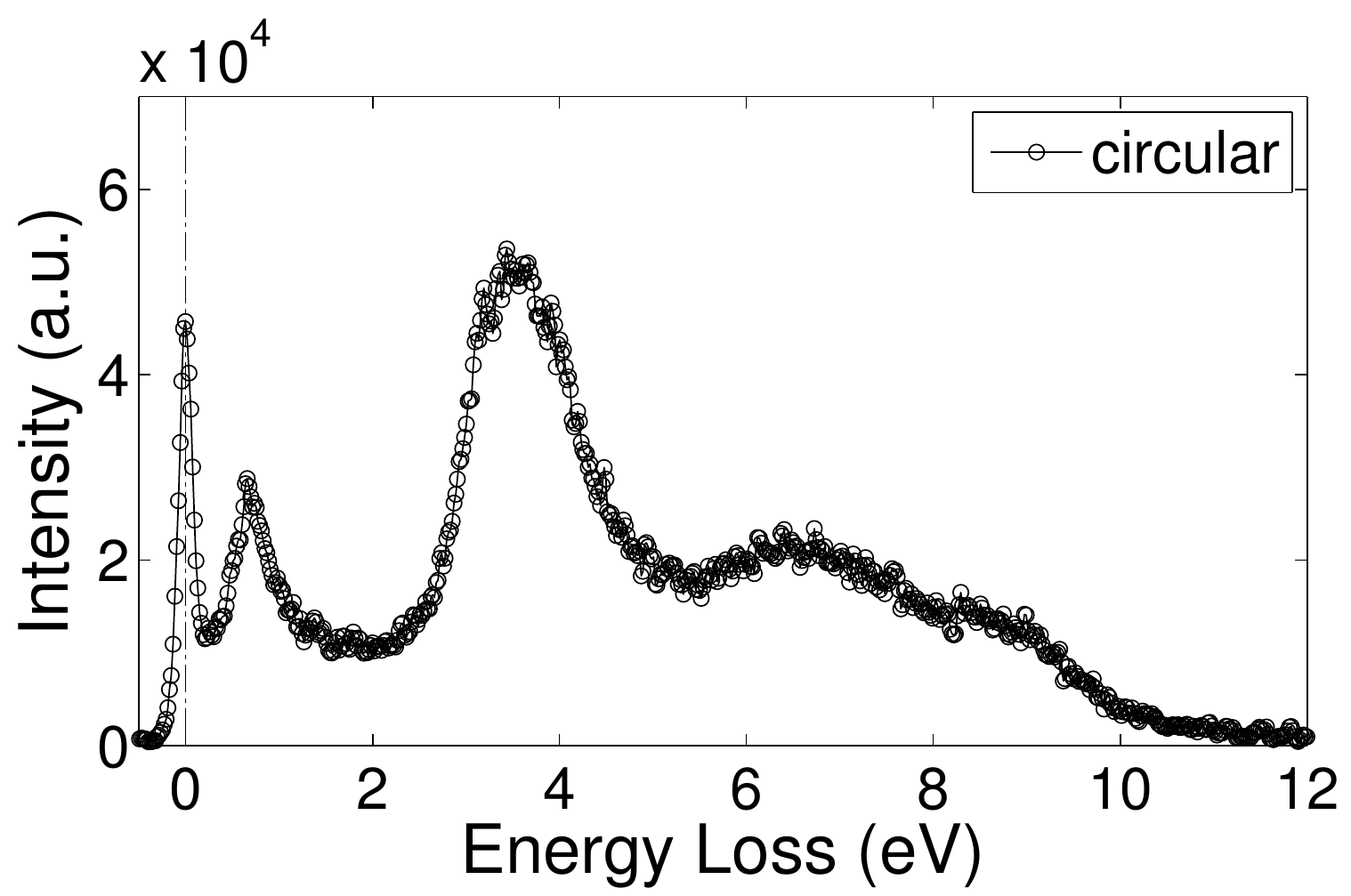}
\end{center}
\caption{RIXS spectra with a circular polarized incident beam in order to unambiguously determine the elastic energy, which was set as 0~eV. The peak at 0.66~eV is excitation of $d$-$d$ transition and higher energy features are dominated by fluorescence.}
\label{fig:circular}
\end{figure}
X-ray polarization and the scattered beam directions \cite{Ament2011}, the intensity of elastic scattering in soft x-ray RIXS measurements can sometimes be much weaker than the inelastic RIXS intensity \cite{DeanReview2015}. In cases like these it can be difficult to unambiguously determine the exact elastic energy in the RIXS spectrum. Therefore for each \mathbi{Q} point measured in the experiments, we first used a circularly polarized incident beam to get appreciable elastic line signal to precisely calibrate the zero energy loss definition on the CCD detector. The result is shown in figure~\ref{fig:circular}. Except the sharp elastic line at zero energy loss, there is a peak at 0.66~eV and multiple features above 2~eV. The peak at 0.66~eV corresponds to the $d$-$d$ transition whose energy is consistent with that observed with Ir $L$-edge RIXS \cite{Kim2012}. Features at higher energy are mainly from the oxygen $K$-edge fluorescence.     

\begin{figure}[!t]
\begin{center}
\includegraphics[width=8 cm]{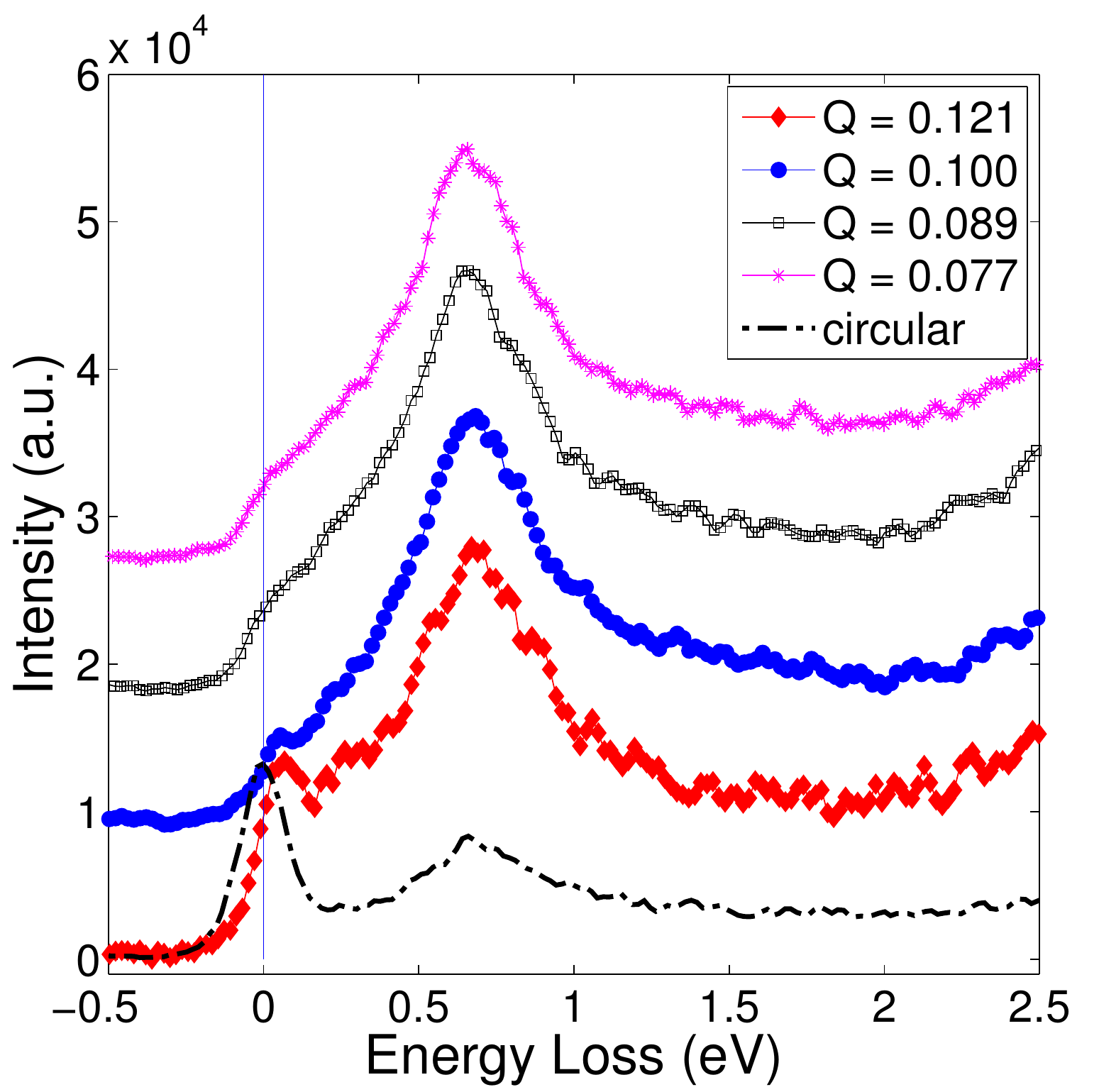}
\end{center}
\caption{\mathbi{Q}-dependence of the RIXS spectra with $\pi$ incident polarization. The dashed line is with circular polarization, as shown in figure \ref{fig:circular}.}
\label{fig:Qdep}
\end{figure}

To better observe the low energy excitations, such as magnons, the overwhelming contribution from the elastic line needs to be minimized. This was done by switching to a $\pi$ polarized incident beam. With the $\pi$ incident polarization, the non-spin-flipped scattering process is strictly forbidden at the scattering angle of $90^\circ$ \cite{Ament2011}. Since our detector was placed close this ($85^\circ$), the elastic line was largely suppressed. Figure~\ref{fig:Qdep} plots RIXS spectra collected at this condition. Indeed the elastic scattering is almost invisible, and low energy excitation peaks appear.  

%\begin{figure}[h]
%\includegraphics[width=0.45\textwidth]{figure3.pdf}

Since the magnetic interactions in Sr$_2$IrO$_4$ are almost completely two dimensional, we index the momentum transfer with only its $H$ component. For $\mathbi{Q} = 0.121$ and $0.100$, the low energy peaks are at 52 and 44~meV respectively, with an error bar of 10~meV (estimated from half of the scanning step size). These energies agree very well with the magnon dispersion measured earlier \cite{Kim2012}. Thus we conclude that the observed low energy peaks are from single magnon excitations. The observed low energy magnon excitation shows distinct \mathbi{Q} dependence: its intensity quickly reduces when the in-plane momentum transfer moves towards $H = 0$. For the other two \mathbi{Q} points closer to $H= 0$, the magnon excitations are barely observable. This behavior is consistent with recent theoretical calculations by B. H. Kim \emph{et al}.\ \cite{KimarXiv2014}. They predict that  oxygen $K$-edge RIXS measurements of single magnons in Sr$_2$IrO$_4$ have a minimum in intensity at $H = 0$, and its size is governed by the buckling of the Ir-O-Ir bond angle. The agreement with this theoretical prediction further confirms that the observed low energy peaks are indeed from single magnon excitations.

In conclusion, we show that single magnon excitations in antiferromagnetically ordered Sr$_2$IrO$_4$ can be probed with RIXS at the $K$-edge of the ligand oxygen atoms. This technique helps to overcome the resolution limitation of $L$-edge RIXS in the hard X-ray region for high $Z$ index transition metal compounds. It has excellent potential to be applied to a wide range of magnetic systems based on heavy elements, and crucially, it is furthermore possible to measure very small sample volumes including thin films and heterostructures, something that until now as been largely limited to cuprate-based systems \cite{Dean2012}.

\section*{Acknowledgments}
We thank Fan Wei and Jeroen van den Brink for helpful discussions. X Liu was supported by the Strategic Priority Research Program of the Chinese Academy of Sciences (Grant No: XDB07020200). The work at Brookhaven National Laboratory was supported by the U.S.\ Department of Energy (DOE), Division of Materials Science, under Contract No.\ DE-SC0012704. J Liu is supported by the Director, Office of Science, Office of Basic Energy Sciences, Materials Sciences and Engineering Division, of the U.S. Department of Energy under Contract No. DE-AC02-05CH11231 through the Quantum Materials program.


\section*{References}
\begin{thebibliography}{20}
\bibitem{Ament2011} Ament L J P \ital 2011 {\it Rev. Mod. Phys.} {\bf 83} 705 
\bibitem{DeanReview2015} Dean M P M 2015 {\it Journal of Magnetism and Magnetic Materials} {\bf 376} 3 
\bibitem{KimReview2012} Kim Y J \ital 2012 {\it Synchrotron Radiation News} {\bf 25-4} 3 
\bibitem{Ament2009} Ament L J P \ital 2009 {\it Phys. Rev. Lett.} {\bf 103} 117003 
\bibitem{Braicovich2010} Braicovich L \ital 2010 {\it Phys. Rev. Lett.} {\bf 104} 077002
\bibitem{Kim2012} Kim J \ital 2012 {\it Phys. Rev. Lett.} {\bf 108} 177003 
\bibitem{Kim2012_327} Kim J \ital 2012 {\it Phys. Rev. Lett.} {\bf 109} 157402 
\bibitem{Yin2013} Yin W G \ital 2013 {\it Phys. Rev. Lett.} {\bf 111} 057202 
\bibitem{Kim2014} Kim J \ital 2014 {\it Nature Communications} {\bf 5} 4453 
\bibitem{LeePRL2013} Lee W S \ital 2013 {\it Phys. Rev. Lett.} {\bf 110} 265502 
\bibitem{SIX} SIX project at the National Syncrotron Light Source II, http://www.bnl.gov/ps/nsls2/beamlines/files/pdf/SIXposter.pdf. 
\bibitem{ERIXS} Upbl07 project at the European Synchrotron Radiation Facility.  
\bibitem{I21} I21 webpage at the Diamond Light Source, http://www.diamond.ac.uk/Home/Beamlines/I21.html/.
\bibitem{VERITAS} Webpage for the VERITAS beamline at Max IV, https://www.maxlab.lu.se/veritas. 
\bibitem{Huang2014} Huang D J and Chang S L 2014 {\it Synchrotron Radiation News} {\bf 27-1} 10 
\bibitem{Chiuzbaian2014} Chiuzb\u{a}ian S G \ital 2014 {\it Review of Scientific Instruments} {\bf 85} 043108 
\bibitem{Serrao2013} Serrao C R \ital 2013 {\it Phys. Rev. B} {\bf 87} 085121
\bibitem{Sala2014} Sala M M \ital 2014 {\it Phys. Rev. B} {\bf 89} 121101 
\bibitem{KimarXiv2014} Kim B H and Brink J V D {\it ArXiv e-prints} 1404.2040 
\bibitem{Dean2012} Dean M P M \ital 2012 {\it Nat. Mater.} {\bf 11} 850 
\end{thebibliography}
\end{document}